\documentstyle[12pt]{article}
\date{}
\def\be{\begin{equation}}
\def\ee{\end{equation}}
\def\bea{\begin{eqnarray}}
\def\eea{\end{eqnarray}}
\def\s{\sigma}

\def\lm{\lambda}
\def\de{\delta}
\def\om{\omega}
\def\pr{\prime}
\title{Instability of Classic Rotational Motion for
Three-String Baryon Model}
\author{G.\,S. Sharov\\
 {\small Tver state university}\\
{\small Tver, 170002, Sadovyj per. 35, Mathem. dep-t, \
E-mail: german.sharov@tversu.ru}}
\begin{document}
\maketitle
\begin{abstract}

The dynamics of baryon string model Y (three-string) is considered with
using the approach that implies defining a classical motion of the system
on the base of given initial position and initial velocities of string points.
The analysis resulted, in particular, in the necessity of regarding more
wide (than it was supposed earlier) class of parametrizations of the world
surface for the adequate description of the three-string model.
This approach let us to ascertain that the classic rotational motions
(flat uniform rotations) for this string configuration are unstable.
Any small asymmetric disturbances grow and inevitably result in
falling a quark into the junction.

\end{abstract}

\section*{Introduction}

In the considered ``three-string" or ``Y" string baryon model [1\,--\,4]
three relativistic strings coming from
three material points (quarks) join at the fourth massless point (junction).
There is three other string baryon models with different topology:
the quark-diquark model $q$-$qq$ \cite{Ko};
the linear configuration $q$-$q$-$q$ \cite{lin}
and the ``triangle" model or $\Delta$-configuration \cite{Tr,PRTr}.

The problem of choosing the most adequate string baryon model among
the four mentioned ones has not been solved yet.
Investigation of this problem from the point of view of the QCD limit
at large interquark distances has not been completed. In particular,
the QCD-motivated baryon Wilson loop operator approach
gives some arguments in favour of the Y-configuration \cite{Isg} or
the ``triangle" model \cite{Corn}.

On the other hand all mentioned string baryon models may be used
for describing orbitally excited baryon states on the Regge trajectories
\cite{Ko,4B}. The rotational motions (planar uniform rotations of the
system) are considered as the orbitally excited baryon states.
The exact classical solutions describing the rotational motions are known
for all these models. For the configurations $q$-$qq$ and $q$-$q$-$q$
the rotating string has the form of a rectilinear segment \cite{Ko,BN}.
For the model ``triangle" this form is the rotating closed
curve consisting of segments of a hypocycloid \cite{Tr,PRTr}.
The form of rotating three-string configuration is three rectilinear
string segments joined in a plane at the angles 120${}^\circ$ \cite{AY,PY}.

When we choose the adequate string baryon configuration we are to take into
account the stability of rotational motions for these systems.
For some models this problem was solved with using numerical analysis
of the disturbed motions close to rotational ones \cite{lin,sttr}.
In particular, the rotational motions of the $q$-$q$-$q$ system with the middle
quark at rest are unstable with respect to centrifugal moving away
of the middle quark \cite{lin}. Any small disturbance results in
a complicated quasiperiodic motion, but the system doesn't transform
into the quark-diquark one \cite{Ko}.
The latter $q$-$qq$ configuration (or the meson string model)
seems to be stable but this question is under investigation.
The simple rotational motions \cite{Tr,PRTr} of the model ``triangle"
are stable for all values of the quark masses $m_i$ and the energy
of the system \cite{sttr}.

The stability problem for the three-string baryon model
hasn't studied yet. In this paper the approach of Refs.~\cite{lin,sttr}
is used for its solving.

In Sect.\,1 the equations of motion with the boundary conditions
for the Y configuration are deduced and the necessity of considering
more wide class of parametizations of the world surface is substantiated.
In Sect.\,2 the solution of the initial-boundary value problem with arbitrary
initial conditions is suggested and the stability of the rotational motions is investigated.

\section{Dynamics of three-string baryon model}

In the three-string baryon model or Y-configuration [1\,--\,4]
three world sheets (swept up by three segments of the relativistic string in
$D$-dimensional Minkowski space)
are parametrized with three different functions $X_i^\mu(\tau_i,\s)$.
It is convenient to use the different notations
$\tau_1$, $\tau_2$, $\tau_3$ for the ``time-like" parameters \cite{Collins}.
But the ``space-like" parameters are denoted here the same symbol $\s$.
These three world sheets are joined along the world line of the junction
that may be set as $\s=0$ for all sheets without loss of generality
(see below).

Under these notations the action of the Y-configuration with masses
$m_i$ of material points at the ends of the strings has the form
\be
S=-\sum_{i=1}^3\int d\tau_i\bigg[\gamma\!\!\int\limits_0^{\s_i(\tau_i)}
\!\!\sqrt{-g_i}\,d\s+m_i\sqrt{V_i^2(\tau_i)}\,\bigg].
\label{S}\ee
Here $\gamma$  is the string tension,
$-g_i=(\dot X_i,X'_i)^2-\dot X_i^2X_i'{}^2$,
$\dot X_i^\mu=\partial_{\tau_i}X_i^\mu$,
$X_i^{\pr\mu}=\partial_\sigma X_i ^\mu$, $(a,b)=a_\mu b^\mu$,
the signature is $(+,-,-,\dots)$,
$V_i^\mu(\tau_i)=\frac d{d\tau}X^\mu(\tau,\s_i(\tau))$ is the tangent
vector to the i-th quark trajectory $\s=\s_i(\tau)$, the speed
of light $c=1$.

Action (\ref{S}) with different $\tau_i$ generalizes the similar
expressions in Refs.~\cite{AY,Collins,PY} (where $m_i=0$)
and in Ref.~\cite{Koshkar}, where the limited class of motions of the system with $m_i\ne0$ is considered.

In the junction of three world sheets $X_i^\mu(\tau_i,\s)$ the
parameters $\tau_i$ are connected in the following general manner:
$$\tau_2=\tau_2(\tau),\quad\tau_3=\tau_3(\tau),\quad\tau_1\equiv\tau.$$
So the condition in the junction takes the form
\be
X_1^\mu\big(\tau,0\big)=X_2^\mu\big(\tau_2(\tau),0\big)=
X_3^\mu\big(\tau_3(\tau),0\big).
\label{junc}\ee

The equations of motion
$$
\frac\partial{\partial\tau_i}\frac{\partial\sqrt{-g_i}}{\partial\dot X_i^\mu}
+\frac\partial{\partial\s}\frac{\partial\sqrt{-g_i}}{\partial X_i^{\pr\mu}}=0,
\qquad i=1,2,3$$
and the boundary conditions for the junction and  the
quark\footnote{The term ``quark" instead of ``material point" is used here
and below for brevity. On the classic level we neglect the spin
and other quantum numbers of quarks.} trajectories may be deduced from action
(\ref{S}).

Using the invariance of action (\ref{S}) with respect to nondegenerate
reparametrizations $\tau_i,\s\;\to\;\tilde\tau_i,\tilde\s$ \cite{BN} on each
world sheet one may set the coordinates in which the orthonormality conditions
\be
\dot X_i^2+X_i^{\pr2}=0,\qquad(\dot X_i,X_i')=0,\qquad i=1,2,3\label{ort}\ee
are satisfied. The junction condition (\ref{junc}) unlike more rigid condition
with $\tau_1=\tau_2=\tau_3$ on the junction line \cite{AY,PY}
let us make these reparametrizations independently on each world sheet.
After this substitution (new coordinates are also denoted $\tau_i,\s$)
the inner equations of the junction line will have a more general form
$\s=\s_{0i}(\tau_i)$ in comparison with the previous one $\s=0$.

Under conditions (\ref{ort}) the equations of motion become linear
\be
\ddot X_i^\mu-X_i^{\pr\pr\mu}=0,\label{eq}\ee
and the boundary conditions in the junction and on the quark
trajectories $\s=\s_i(\tau_i)$ take the following form:
\bea
&&\;\sum_{i=1}^3\Big[X_i^{\pr\mu}\big(\tau_i(\tau),\s_{0i}(\tau_i)\big)
+\s_{0i}'(\tau_i)\,\dot X_i^\mu\big(\tau_i,\s_{0i}(\tau_i)\big)\Big]
\tau'_i(\tau)=0,\qquad\label{qy}\\
&&m_i\frac d{d\tau_i }U_i^\mu(\tau_i)+\gamma \big[X_i^{\pr\mu}+\s_i'(\tau_i)
\,\dot X_i^\mu\big]\Big|_{\s=\s_i(\tau_i)}=0,\quad i=1,2,3. \qquad
\label{qq}\eea
Here $U^\mu_i(\tau_i)=V^\mu_i(\tau_i)\big/\sqrt{V^2_i(\tau_i)}$
is the unit velocity vector of $i$-th quark.

The reparametrizations
\be
\tilde\tau_i\pm\tilde\s=f_{i\pm}(\tau_i\pm\s),\qquad i=1,2,3
\label{rep}\ee
($f_{i\pm}$ are arbitrary smooth monotone functions \cite{BN}) keep
invariance of conditions (\ref{ort}), equations (\ref{eq}) and boundary
conditions (\ref{qy}), (\ref{qq}). Choosing the functions $f_{i\pm}$
we can fix the equation of the junction line and the quark trajectories
on all world sheets in the form
\be
\s_{0i}(\tau_i)=0,\quad\s_i(\tau)=\pi,\qquad 0\le\s\le\pi,\qquad i=1,2,3,
\label{fiq}\ee
On each world sheet independently one can obtain the condition
``$\s=0$ is the junction trajectory" by choosing the function $f_{i+}(\xi)$
in Eq.~(\ref{rep}) and keeping identical $f_{i-}(\xi)=\xi$.
Then one can obtain the equalities $\s_i=\pi$ through the next substitution
(\ref{rep}) with $f_{i+}=f_{i-}$ (it keeps invariance of the equation $\s=0$).

In this paper the parametrization satisfying the conditions
(\ref{fiq}) and (\ref{ort}) is used. The alnernative approach is possible.
It implies introducing the condition $\tau_2(\tau)=\tau_3(\tau)=\tau$
on the junction trajectory (\ref{junc}) in conjunction with
the condition $\s_{0i}=0$ (or $\s_{0i}={}$const).
But under such circumstances the functions $\s_i(\tau)$ in the quark
trajectories are not equal to constants in general.

If under orthonormal gauge (\ref{ort}) we demand satisfying
as conditions (\ref{fiq}), as the equalities $\tau_1=\tau_2=\tau_3$
on the junction line (\ref{junc}), then we actually restrict the
class of motions of the system which the model describes.
In other words, not all physically possible motions satisfy
the above mentioned conditions. Such a situation without explicit
indication takes place, in particular, in Ref.~\cite{PY}.

For the proof of these statements note that after introducing the
restrictions (\ref{ort}) and (\ref{fiq}) we shall have the class
of reparametrizations (\ref{rep}) keeping these conditions \cite{PeSh}.
In them the functions $f_{i+}(\xi)=f_{i-}(\xi)=f_i(\xi)$ have the form
\be
f_i(\xi+2\pi)=f_i(\xi)+2\pi,\qquad f'_i(\xi)>0.
\label{MP}\ee
The functions from the class (\ref{MP}) may be represented in the form
\cite{PeSh} $f(\xi)=\xi+\phi(\xi)$, $\phi(\xi+2\pi)=\phi(\xi)$,
$\phi'(\xi)>-1$. They have the following (directly verifiable) properties:
if $f(\xi)$ and $g(\xi)$ satisfy the conditions (\ref{MP}), then
the inverse function $f^{-1}(\xi)$ and the superposition
$f\big(g(\xi)\big)$ also satisfy (\ref{MP}).

Making transformations (\ref{rep}), (\ref{MP}) we demand satisfying
the equalities $\tilde\tau_2=\tilde\tau_3=\tilde\tau_1$ on the junction
line $\s=0$, that results in the relations
$$f_2\big(\tau_2(\tau)\big)=f_3\big(\tau_3(\tau)\big)=
f_1(\tau)\;\;\Longrightarrow\;\;\tau_2(\tau)=f_2^{-1}\big(f_1(\tau)\big),
\quad\tau_3(\tau)=f_3^{-1}\big(f_1(\tau)\big).$$
Hence, one may obtain the equalities $\tilde\tau_2=\tilde\tau_3=\tilde\tau_1$
for all $\tilde\tau$ under conditions (\ref{ort}) and (\ref{fiq}) only if
the functions $\tau_2(\tau)$ and $\tau_3(\tau)$
{\it satisfy the conditions} (\ref{MP}). It will be shown in Sect.~2
that the majority of physical motions of the three-string do not
satisfy this restriction (see, for example, Fig.~1f).

For describing an arbitrary motion of the three-string in the suggested
approach the unknown functions $\tau_i(\tau)$ are determined
from dynamic equations, in particular, with solving the initial-boundary
value problem.

\section{Initial-boundary value problem
and stability of the rotational motions}

The initial-boundary value problem implies obtaining the motion of the
three-string on the base of two given initial conditions:
an initial position of the system in Minkowski space and initial
velocities of string points. In other words, we are to determine
the solution of Eq.~(\ref{eq}) $X_i^\mu(\tau_i,\s)$, sufficiently smooth \cite{lin,BaSh} and satisfying the orthonormality (\ref{ort}), boundary
(\ref{junc}), (\ref{qy}), (\ref{qq}) and initial conditions.

An initial position of the three-string can be given as three joined curves
in Minkowski space with parametrizations
$$x^\mu=\rho^\mu_i(\lm),\qquad\lm\in[0,\lm_i],\qquad
\rho^\mu_1(0)=\rho^\mu_2(0)=\rho^\mu_3(0), \qquad{\rho'_i}^2<0.$$
Initial velocities on the initial curves can be given as a time-like
vectors $\,v_i^\mu(\lm)$, $\lm\in[0,\lm_i]$, $v_i^\mu(\lm)$ may be
multiplied by an arbitrary scalar function $\chi(\lm)>0$.
The condition $v_i^\mu(0)=v_j^\mu(0)\cdot{}$const in the junction
is fulfilled.

To solve the problem we set parametrically the initial curves
on the world sheets
$$\tau_i=\tau_i^*(\lm),\qquad\s=\s_i^*(\lm),\qquad\lm\in[0,\lm_i],$$
and use the following general form for the initial position of the three
segments of the Y configuration \cite{BaSh}:
\be
X_i^\mu\bigl(\tau_i^*(\lm),\s_i^*(\lm)\bigr)=\rho_i^\mu(\lm),
\qquad\lm\in[0,\lm_i],\quad i=1,2,3.
\label{X=r}\ee
Here $|\tau_i^{*\pr}|<\s_i^{*\pr}$, $\tau_i^*(0)=\s_i^*(0)=0$,
$\s_i^*(\lm_i)=\pi$.
There is the freedom in choosing the functions $\tau_i^*(\lm)$, $\s_i^*(\lm)$
connected with the free choice of the functions $f_i$ in Eqs.~(\ref{rep})
satisfying conditions (\ref{MP}).

Let us consider the general solutions of Eq.~(\ref{eq}) on the world sheets
\be
X_i^\mu(\tau_i,\s)=\frac1{2}\bigl[\Psi^\mu_{i+}(\tau_i+\s)+
\Psi^\mu_{i-}(\tau_i-\s)\bigr],\qquad i=1,2,3.
\label{soly}\ee
The derivatives of functions $\Psi^\mu_{i\pm}$ satisfy the isotropy condition
\be
\Psi^{\pr2}_{i\pm}(\tau)=0
\label{iso}\ee
resulting from the orthonormality conditions (\ref{ort}).

Using the formulas \cite{BaSh}
\be
\frac d{d\lm}\Psi^\mu_{i\pm}\bigl(\tau_i^*(\lm)\pm\s_i^*(\lm)\bigr)=
\big[1\pm(v_i,\rho'_i)\,Q_i\big]\,\rho_i^{\pr\mu}(\lm)\mp
Q_i\rho_i^{\pr2}v_i^\mu(\lm),
\label{psi}\ee
where $\,Q_i(\lm)=\big[(v_i,\rho'_i)^2-v_i^2\rho_i^{\pr2}\big]^{-1/2}$,
we can determine from the initial data the function $\Psi^\mu_{i+}$
in the initial segment $\bigl[0,\tau_i^*(\lm_i)+\pi\bigr]$ and the function
$\Psi^\mu_{i-}$ in the segment $\bigl[\tau_i^*(\lm_i)-\pi,0\bigr]$.
The constants of integration are fixed from the initial condition
(\ref{X=r}).

The functions $\Psi^\mu_{i\pm}$ are to be continued beyond the initial segments
with the help of the boundary conditions (\ref{junc}), (\ref{qy}), (\ref{qq}).
In particular, the conditions on the quark trajectories (\ref{qq})
after substituting in them Eq.~(\ref{soly}) with taking into account
Eqs.~(\ref{fiq}) are reduced to \cite{lin,BaSh}
\bea
&\displaystyle
\frac{dU^\mu_i}{d\tau_i}=\frac\gamma{m_i}\big[\de^\mu_\nu-U^\mu_i(\tau_i)\,
U_{i\nu}(\tau_i)\big]\Psi^{\pr\nu}_{i-}(\tau_i-\pi),&\label{Uy}\\
&\Psi^{\pr\mu}_{i+}(\tau_i+\pi)=\Psi^{\pr\mu}_{i-}(\tau_i-\pi)-
2m_i\gamma^{-1}U^{\pr\mu}_i(\tau_i).\label{psy}&
\eea
The systems of equations (\ref{Uy}) need the initial conditions
$U^\mu_i\big(\tau^*(\lm_i)\big)=v_i^\mu(\lm_i)\big/\sqrt{v_i^2(\lm_i)}$,
$i=1,2,3$.

Substituting Eq.~(\ref{soly}) into the boundary conditions in the junction
(\ref{junc}) and (\ref{qy}) we express the function $\Psi^{\pr\mu}_{i-}(\tau_i)$
through $\Psi^{\pr\mu}_{i+}(\tau_i)$:
\be
\left(\begin{array}{c}
\Psi^{\pr\mu}_{1-}(\tau) \\
\tau'_2\Psi^{\pr\mu}_{2-}\big(\tau_2(\tau)\big) \\
\tau'_3\Psi^{\pr\mu}_{3-}\big(\tau_3(\tau)\big)\end{array}\right)=\frac13
\left(\begin{array}{ccc} -1 & 2 & 2\\  2 & -1 & 2\\ 2 & 2 & -1
\end{array}\right)
\left(\begin{array}{c} \Psi^{\pr\mu}_{1+}(\tau)\\
\tau'_2\Psi^{\pr\mu}_{2+}(\tau_2) \\ \tau'_3\Psi^{\pr\mu}_{3+}(\tau_3)
\end{array}\right).
\label{jps}\ee

Eqs.~(\ref{jps}), (\ref{psy}) and (\ref{Uy}) let us infinitely continue
the functions $\Psi^\mu_{i\pm}$ outside the initial segments if the
functions $\tau_2(\tau)$ and $\tau_3(\tau)$ are known. They will be found with
using Eqs.~(\ref{iso}) and the relations
\be
[\tau'_i(\tau)]^2\big(\Psi'_{i+}(\tau_i),\Psi'_{i-}(\tau_i)\big)=
\big(\Psi'_{1+}(\tau),\Psi'_{1-}(\tau)\big),\qquad i=2,3,
\label{cont}\ee
obtained from Eqs.~(\ref{junc}) and (\ref{soly}).
Only under the conditions (\ref{cont}) the isotropy of vector-functions
$\Psi^{\pr\mu}_{i+}$ in the r.h.s. of Eqs.~(\ref{jps}) results in the isotropy
of $\Psi^{\pr\mu}_{i-}$ in the l.h.s.

Substituting $\Psi^{\pr\mu}_{i-}$ from Eq.~(\ref{jps}) into Eq.~(\ref{cont})
we obtain the relations for calculating the functions $\tau_2(\tau)$ and
$\tau_3(\tau)$:
\be
\tau'_2(\tau)=\frac{\big(\Psi'_{1+}(\tau),\Psi'_{3+}(\tau_3)\big)}
{\big(\Psi'_{2+}(\tau_2),\Psi'_{3+}(\tau_3)\big)},\qquad
\tau'_3(\tau)=\frac{\big(\Psi'_{1+}(\tau),\Psi'_{2+}(\tau_2)\big)}
{\big(\Psi'_{2+}(\tau_2),\Psi'_{3+}(\tau_3)\big)}.
\label{taus}\ee
Here the functions $\Psi^{\pr\mu}_{i+}$ are taken from Eqs.~(\ref{psi}) and
(\ref{psy}) during solving the initial-boundary value problem.

Eqs.~(\ref{Uy}), (\ref{psy}), (\ref{jps}) and (\ref{taus}) form the closed
system for infinite continuation of the functions $\Psi^\mu_{i\pm}$.
Determining $X_i^\mu$ from Eq.~(\ref{soly}) we solve (numerically, in general)
the initial-boundary value problem for the three-string model.

The described method of solving this problem is used here for investigating
the stability of the rotational motions. For this purpose we are to consider
the initial-boundary value problem with disturbed (variously) initial
conditions $\rho_i^\mu(\lm)$ and $v_i^\mu(\lm)$.

As was mentioned above the rotational motion of the Y configuration is uniform
rotating of three rectilinear string segments with lengths $R_i$ joined
in a plane at the angles 120${}^\circ$ \cite{AY,PY}. The lengths $R_i$ are
connected with the angular velocity $\om$ by the relation \cite{Ko,lin,4B}
\be
R_i\om^2(R_i+m_i/\gamma)=1.
\label{Rom}\ee

This rotational motion may be obtained by solving the initial-boundary value
problem with appropriate initial position $\rho_i^\mu(\lm)$ (the rectilinear
segments with lengths $R_i$) and velocities
\be
v_i^\mu(\lm)=\{1; \vec v_i(\lm)\},\qquad
\vec v_i(\lm)=\big[\vec\om\times\vec\rho_i(\lm)\big]+\de\vec v_i(\lm)
\label{vel}\ee
with the disturbance $\de\vec v_i=0$.

To test the stability of this motion we consider the initial conditions with
small disturbances $\de\rho_i^\mu(\lm)$ or $\de v_i^\mu(\lm)$.

The typical example of a slightly disturbed quasirotational motion of the
three-string with $m_1=m_2=m_3=1$, $\gamma=1$ is represented in Fig.\,1.
Here all $R_i=0.3$, $\om$ satisfy the condition (\ref{Rom}). The initial
velocities have the form (\ref{vel}) where only $x$-component of the
velocity disturbance for the first string segment is non-zero:
$\de v_1^1(\lm)=0.05\lm$ ($0<\lm<\lm_1=1$).

In Figs.\,1a\,--\,e the positions of the system in $xy$-plane (sections
$t={}$const of the world surface) are shown. For saving in space the axes are
sometimes omitted. They are numbered in order of
increasing $t$ with the step in time $\Delta t=0.15$ and these numbers are
near the position of the first quark marked by the small square. The point here
and below marks the positions of the 2-nd quark and the circle --- of the 3-rd
quark.

The picture of motion is qualitatively identical for any small asymmetric
disturbance $\de\rho_i^\mu(\lm)$ or $\de v_i^\mu(\lm)$. Starting from some
point in time (position 5 in Fig.\,1a) the junction begins to move. During
this complicated motion the distance between the junction and the rotational
center increases and the lengths of the string segments vary quasiperiodically
(Figs.\,1b\,--\,d) unless one of these lengths become equal to zero, i.e.
a quark falls into the junction. In Fig.\,1e this situation takes place after
the shown position 81 (there is some missed time interval between Figs.\,1d
and 1e).

Thus the numerical modeling in this and other numerous examples shows that
rotational motions of the three-string are {\it unstable}. The evolution of
the instability is slow at the first stage if the disturbance is small, but the
middle and final stages are rather similar to the motion in Fig.\,1a\,--\,e.

In Fig.\,1f the dependencies $\tau_2(\tau)$ (the full line) and $\tau_3(\tau)$
(the dotted line) for this motion are represented. The speed of the ``time
flow" $\tau_i$ increases with diminishing the correspondent string segment.
The horizontal asymptotes of the curves in Fig.\,1f are connected with
vanishing the length of the first string segment. As we see, the functions
$\tau_2(\tau)$ and $\tau_3(\tau)$ do not satisfy the periodicity
conditions (\ref{MP}). This fact does not allow describing this motion
in the frameworks of the parametrization \cite{AY,PY} with
$\tau_1=\tau_2=\tau_3$.

Any slightly disturbed rotational motion of the three-string is finished with
falling one of the quarks into the junction. Considering the further classical
evolution of this system\footnote{This problem is pure theoretical one:
classical passing of the material point through the junction will unlikely be
the same after the quantization or developing a more general QCD-based theory.}
in the framework of action (\ref{S}) we must take into account that vanishing
the $i$-th string segment is simultaneous with the becoming infinite the
corresponding ``time" ($\tau_i\to\infty$). This is not only ``bad
parametrization" but the geometry of the system changes: the three-string
transforms into the linear $q$-$q$-$q$ configuration after merging a quark
with the junction. The lifetime of this ``$q$-$q$-$q$ stage" is finite and
non-zero so the material point with the mass $m_i$ moving at a speed $v<1$
can not slip through the junction instantaneously. Otherwise under three
non-compensated tension forces the massless junction will begin to move
at the speed of light.

To illustrate this process in Fig.\,2 the motion of the three-string with different
masses at the ends
$m_1=1$, $m_2=2$, $m_3=3$, $\gamma=1$ is represented (the notations are
the same, $\Delta t=0.125$). It is close to the
rotational one: the initial velocities satisfy the relation (\ref{vel})
with $\de v_1^1(\lm)=0$, the angular velocity $\om\simeq1.6$ and the
different lengths $R_1=0.3$, $R_3\simeq0.125$ are connected by Eqs.~(\ref{Rom}).
But the assigned value $R_2=0.22$ does not satisfy (\ref{Rom}) (that
gives $R_2\simeq0.179$) so this difference plays a role of the disturbance
for the motion in Fig.\,2. After the position 31 in Fig.\,2c the 3-rd quark
falls into the junction (this is more probable for the heaviest quark if their
masses are different).

The 3-rd material point after falling merges with the junction. They move
together (Fig.\,2d) unless two other string segments form at the junction
the angle 120${}^\circ$. The waves from the point of merging spread along
the strings so the motion become more complicated.

The above described behavior of slightly disturbed rotational motions
takes place also for the massless ($m_i=0$) three-string model \cite{AY,PY}.
To solve the initial-boundary value problem for this system with using the
suggested approach, we are to substitute Eqs.~(\ref{Uy}), (\ref{psy})
for the equation
$$\Psi^{\pr\mu}_{i+}(\tau_i+\pi)=\Psi^{\pr\mu}_{i-}(\tau_i-\pi).$$
In Fig.~3 the motion of this configuration is represented with the same
notations. Here in the initial conditions all $R_i=1$, $\om=1$, $\Delta t=0.2$
the velocities
have the form (\ref{vel}) with the small disturbance $|\de v_1^1(\lm)|<0.01$
(note that the initial velocities of the strings' ends are equal to
the speed of light $v=1$). The evolution of the instability is connected with
moving the junction away from the center is shown in Fig.~3a\,--\,d.
In Fig.\,3e the derivatives $\tau'_2(\tau)$, $\tau'_3(\tau)$ for this motion
are represented. The functions $\tau_i(\tau)$ do not satisfy the condition
(\ref{MP}) and such a situation takes place in general.

\section*{Conclusion}

In this paper method of solving the initial-boundary value problem with
arbitrary initial conditions for the string baryon configuration Y
(three-string) is suggested. This approach was used for numerical investigation
of the stability problem for the classic rotational motions of this system.
It was shown that these motions for all values of the parameters $m_i$, $\om$
are {\it unstable}: any small asymmetric disturbances grow with growing time.
The evolution of this instability has some universal features and results
in merging one of the quarks with the junction.

This fact does not totally close the three-string model and, in particular,
its applications for describing the baryonic Regge trajectories. The majority
of orbitally excited baryon states are resonances so the stability of
corresponding classical motions is not necessary for modeling this states.
None the less the results obtained in this paper for the Y configuration
in comparison with other string baryon models\footnote{Remind that the simple
rotational states of the string model ``triangle" are stable \cite{sttr}.}
are important for choosing the most adequate string baryon both for describing
 the baryon states on Regge trajectories
\cite{4B} and for development more perfect QCD-based baryon models.

\begin{figure}[b]
\caption{The slightly disturbed rotational motion of the three-string
with equal masses $m_1=m_2=m_3=1$.}
\end{figure}
\begin{figure}[b]
\caption{The quasirotational motion of the three-string
with masses $m_1=1$, $m_2=0$, $m_3=1$.}
\end{figure}
\begin{figure}[b]
\caption{The slightly disturbed rotational motion of the massless
three-string ($m_1=m_2=m_3=0$).}
\end{figure}

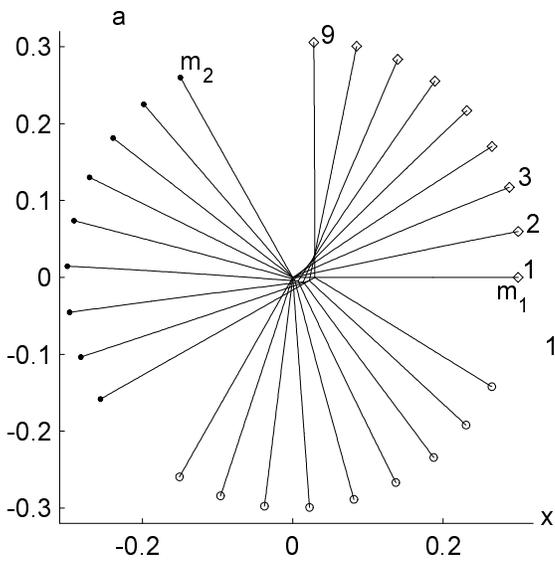
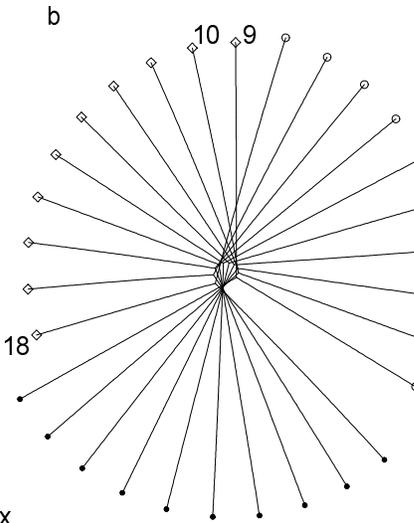
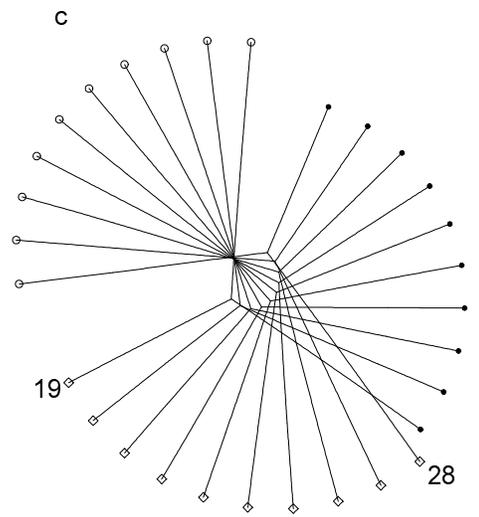
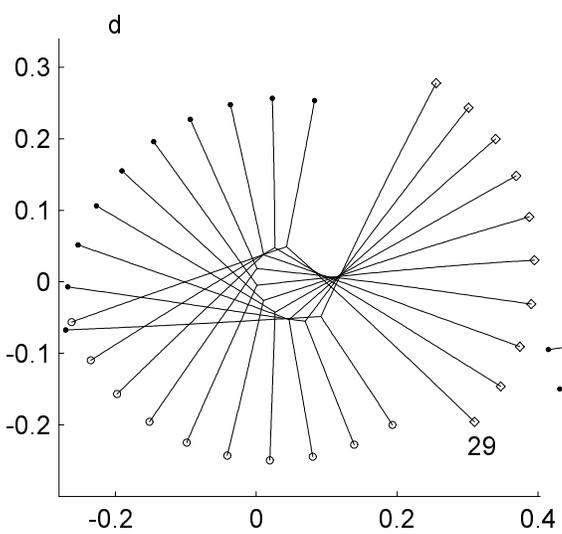
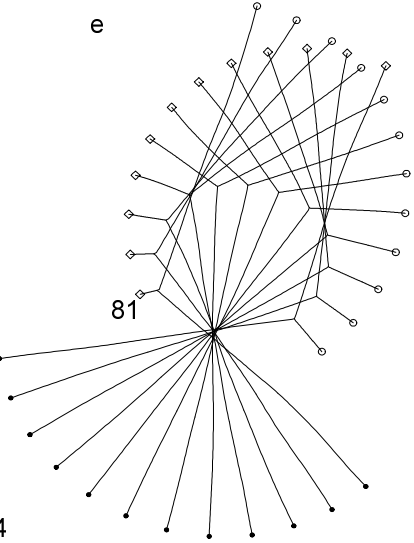
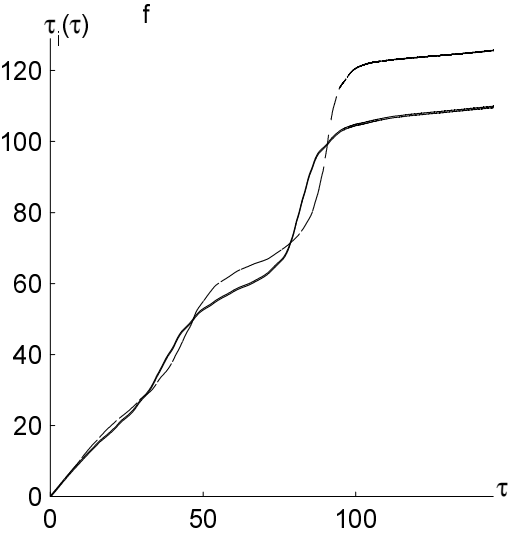
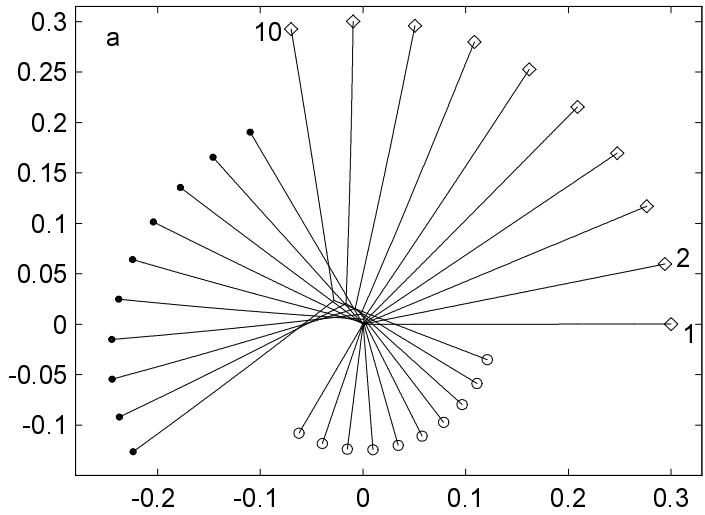
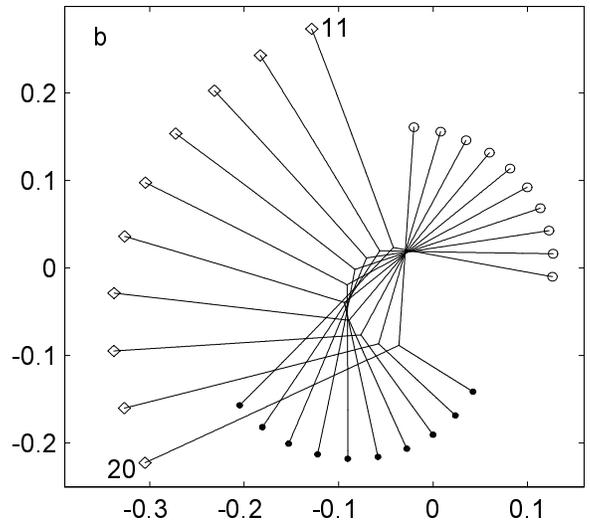
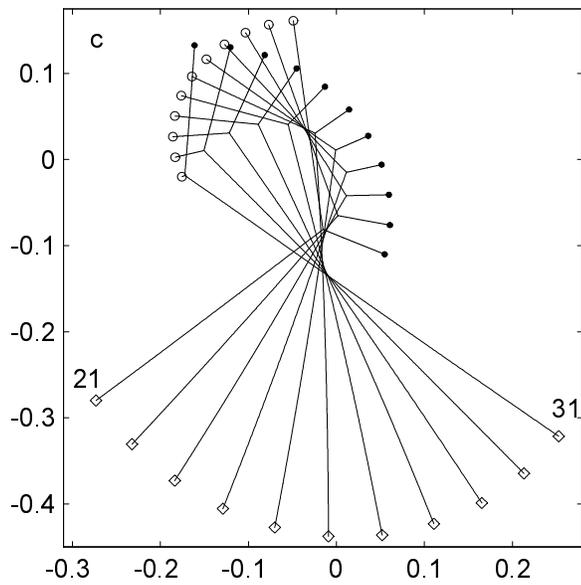
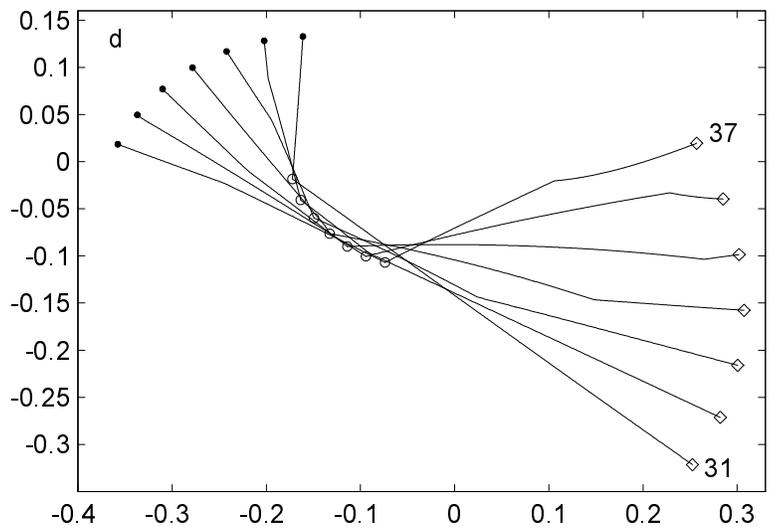
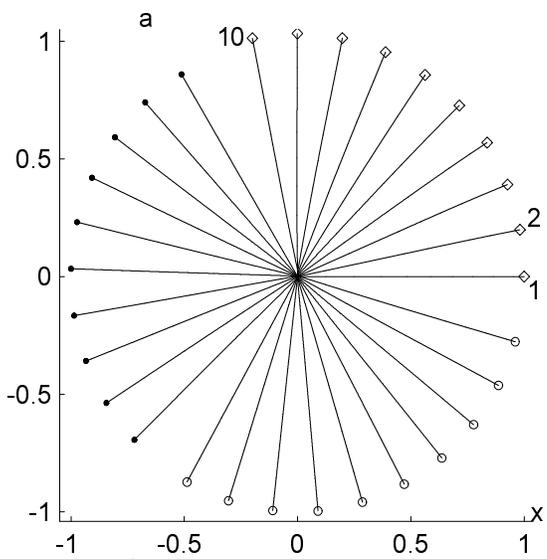
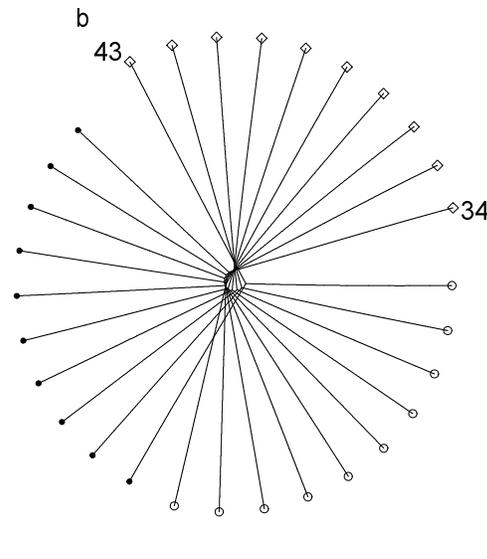
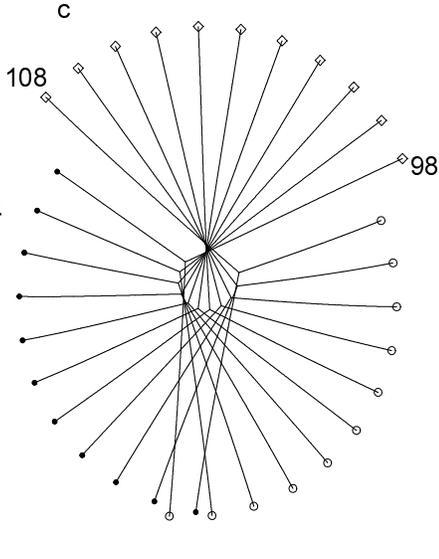
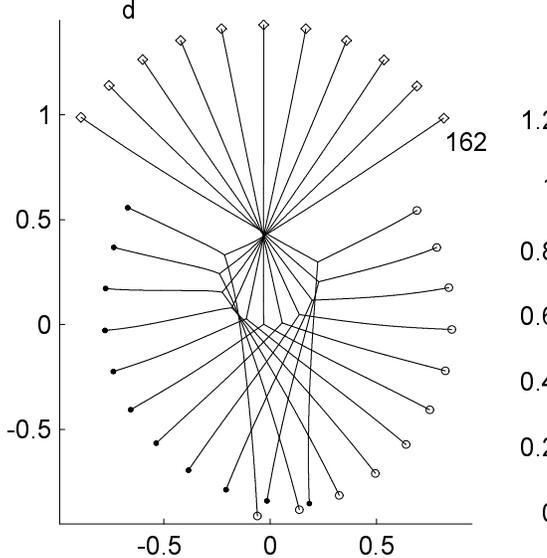
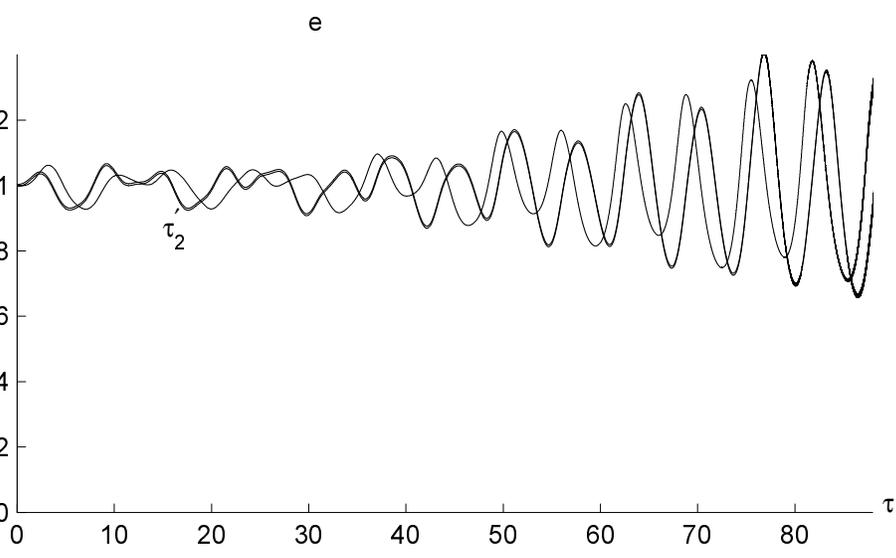
\end{document}